\documentstyle[12pt,psfig]{article}
\begin{document}
\date{June 22, 1998}
\def\Nat{\rm l\!N}
\def\R{\rm l\!R\,}
\def\compton{\overline{\phantom{\overline{.\!.}}}\!\!\!\lambda}
\title{Magnetic Permeability  in\\ Constrained Fermionic Vacuum}
\author{M. V. Cougo-Pinto\thanks{e-mail: marcus@if.ufrj.br}, 
C. Farina\thanks{e-mail: farina@if.ufrj.br},
A. Tort\thanks{e-mail: tort@if.ufrj.br}\\
\\Instituto de F\'\i sica, Universidade Federal do Rio de 
Janeiro\\
CP 68528, Rio de Janeiro, RJ 21945-970, Brazil\\
\\
J. Rafelski\thanks{e-mail: Rafelski@Physics.Arizona.Edu}\\
\\
Department of Physics, University of Arizona\\
Tucson, AZ 85721, USA\\
\\
}
\maketitle
\begin{abstract}
We obtain  using Schwinger's proper time approach 
the Casimir-Euler-Heisenberg effective action of 
fermion fluctuations for the case of an applied magnetic field. 
We implement here the compactification of
one space dimension into a circle through anti-periodic 
boundary condition. Aside of higher order 
non-linear field effects we identify a novel contribution to the 
vacuum permeability. These contributions are exceedingly small 
for normal electromagnetism due to the smallness 
of the electron Compton wavelength compared to
the size of the compactified dimension, if we take the latter
as the typical size of laboratory cavities, but their 
presence is thought provoking, 
also considering the context of strong interactions.
\end{abstract}\vfill
%
There is a major ongoing effort \cite{Iac79,Can91,Bak97} to measure the 
Euler-Heisenberg (EH) effect  \cite{HEW,Sch51}. As it is well understood, 
the EH-effective  Lagrangian arises from
deformation of fermion-antifermion pair fluctuations caused by an 
applied strong (classical = absence of virtual photon diagrams)
electromagnetic field. Since renormalization defines the
electric charge in an Abelian theory in the long wavelength 
limit, the field dependent terms in the effective action
after renormalization must be of higher order than quadratic 
and thus introduce non-linear effects in the electromagnetic field.
A (coherent) light wave within a field-filled volume 
experiences matter-like scattering effects from the field-polarized 
vacuum\cite{Urban}. The hope and expectation is that the birefringence 
effect \cite{BB70} can be experimentally observed 
in near future. 
\eject 
There are several
reasons that make the search for such a macroscopic confirmation 
of vacuum fluctuation effects worthwhile: \\
i) the experiment constitutes a test of quantum electrodynamics in
the far infrared domain  $q\to 0$ in which limit the usually dominant
lowest order vacuum polarization effect vanishes exactly
in consequence of charge renormalization. Thus  one probes
the interaction  of a light wave with the external fields
at the level ${\cal O}(\alpha^2)$;\\ 
ii) there is hope that
when the experimental techniques are refined, certain otherwise 
invisible higher order effects, such as is the interference with
quantum chromodynamic vacuum structure could become observable. 
An effect is in principle possible since quarks are carriers 
of both Maxwell and strong charge \cite{Old}. 
\par
The properties of the (relativistic) quantum 
vacuum are also influenced by boundary conditions constraints, an 
effect  generally known as the Casimir 
effect \cite{Cas48} first studied  for 
the case of two uncharged conductive parallel plates 
restraining the photon fluctuations, causing 
an observable attractive force. 
The effect of constrains on the quantum vacuum 
of a massive field is usually significant when the dimension of the 
support region is comparable to the Compton wavelength 
$\compton=\hbar c/m$ of the field excitations. 
The Casimir effect is thus negligible 
for electrons in nano-cavities. On the other hand, 
in QCD the hadronic confinement region and thus presumably also
the quark field fluctuation region is smaller than $\compton_q$ and
we should expect significant interference between
Euler-Heisenberg and Casimir effects for quark fluctuations.
\par
In view of this observation it is necessary 
to study what new physical phenomena could
arise when the effective action is evaluated for a space-time
region subject to the combined effect of an applied field 
and boundary conditions for the charged fermion fluctuations, 
amalgamating the Euler-Heisenberg and Casimir effects in
a Euler-Heisenberg-Casimir (HEC) effective action.
We are not interested here in the enhancement of the 
Casimir effect caused by the external field \cite{B2}.
Our objective is to understand 
the modifications of the 
vacuum properties, specifically here magnetic 
permeability induced by the Casimir effect.
\par
In other words, we would like to derive the vacuum polarization
effect in the infrared limit $q\to 0$ for finite $ma$, where 
$m$ is the mass of the fermion and $a$ is a length related to the
boundary condition for the Casimir effect. 
In several aspects our interests parallels the study of the cavity
Casimir effect \cite{Scharnhorst-Barton} as well as the consideration 
of radiative corrections to the Casimir effect recently 
obtained by Kong and Ravndal \cite{KR97}. Unlike these efforts
to understand fluctuations of a confined quantum electromagnetic vacuum
our study deals with a confined quantum fermionic vacuum in the presence 
of an (external) classical electromagnetic field. While our work is
carried out for the Abelian Maxwell external fields, it is a study case 
for the physics applicable to the non-Abelian strong interactions, where 
the magnitude of the expected effects is significant. 
\par
To obtain the  interference of the Casimir effect with the external field 
we impose anti-periodic boundary condition on the fermionic field in the 
z-direction from $z=-a$ to $z=a$. This boundary condition corresponds
topologically to a confinement of twisted spinors into a circle ${\rm S}^1$ 
(of radius $a/\pi$) resulting from a compactification of the 
z-axis dimension \cite{Ford80}; 
this choice of twisted spinors avoids the problem of non-causal 
propagation that occurs with the untwisted spinors 
of the periodic boundary condition \cite{Ford80}. The 
anti-periodic boundary condition gives rise to a Casimir effect as
well as the boundary condition of confinement between impermeable 
plates (see the reviews in \cite{Cas48} and for the bosonic case also
\cite{AmbjornWolfram83}) and avoids the mathematical complexity of the 
latter (the $\gamma$ matrices dependent MIT boundary condition \cite{MIT}).
The xy-planes we take as large squares with side $\ell$ and the limit
$\ell\rightarrow\infty$ can be taken at the end of the calculations.
\par
We will consider the case of a constant uniform magnetic field ${\bf B}$
perpendicular to the xy-plane  applied in the vacuum of the 
anti-periodic fermion field. These 
choices for the geometry, the boundary condition and the external 
field are intended to simplify the formalism, in order for us to 
concentrate on the physical effects. Once their basic features are 
understood the path is open to consider more complicated situations,
in particular more complicated geometries and boundary conditions
\cite{Milton83}.
\par 
We note that a simple dimensional and symmetry consideration leads to the 
effective action in lowest order in field
strength in the form
\begin{eqnarray}\nonumber
{\cal L}_{\rm eff}&=&\Pi(ma)\frac{\vec E^{\,2}-\vec B^{\,2}}{2}
+Q(ma) \vec E \cdot \vec B 
+ F_1(ma)\frac{1}{2}( \vec E\cdot\vec n)^2-\\
&-&F_2(ma)\frac{1}{2}( \vec B\cdot\vec n)^2
+F_3(ma)\vec B\cdot \vec n\,\,\vec E\cdot\vec n 
 + {\cal O}(E^4,B^4,E^2B^2)\,,\label{form}
\end{eqnarray}
where $\vec n$ is the normal vector of the xy-plane. We note that 
the terms odd in $\vec n$ cannot occur,
since there is no sense of orientation introduced by the boundary
condition: there are as many fluctuations moving to the `right',
as there are moving to the `left'. In consequence terms odd 
in $\vec n$ cannot occur, which along with particle-antiparticle
symmetry (Furry theorem) assures that the effective action
is an even function in the electromagnetic fields. 
\par
Given the breaking of the symmetry by the boundary condition 
we expect the vacuum state to be birefringent \cite{BB70}. 
For the term $F_3(am)$ to induce the same
magnitude effect as the original Euler-Heisenberg 
birefringence  \cite{Iac79} we must have:
\begin{equation}\label{estimate}
F_3^{\rm equiv}\simeq \vert B\vert ^2 7 A\,,\qquad 
A=\frac{2\alpha^2}{45}\frac{(\hbar/mc)^3}{mc^{2}}=
0.265 \frac{\mbox{fm}^3}{\mbox{MeV}}\,.
\end{equation}
A is the coefficient of the first non vanishing EH term, see
also Eq.\,(\ref{HEserie}) below.
The energy density $B^2/2$ of a 1 Tesla field is
$0.321\times 10^{-18}\,\mbox{keV}/\mbox{fm}^3$.
For a 5 Tesla field we thus find $F_3^{\rm equiv}(ma)
\simeq 3\times10^{-17}$, a small number indeed. This clearly illustrates 
the difficulty of the experimental effort which must reduce 
birefringence due to matter to below this contribution. A different 
view at the smallness of the Maxwell theory effects arises recalling
that the fields are measured in units of the `critical' field,
here $B_{\rm cr}=m^2c^3/e\hbar=4.414\ 10^{9}\,{\rm T}$\,.
\par
The evaluation of all the five factor functions in Eq.\,(\ref{form}), 
allowing for the strong interaction structure of the vacuum,  is a
formidable task which one should undertake only upon
confirmation that there are interesting physical
properties waiting to be discovered. We can explore
one interesting aspect in a relatively ease way  and thus 
motivate further study  of this complex subject matter. 
It turns out that the QED-case $\vec n \parallel \vec
B\,,$ with $\vec E=0$ is easily analytically soluble using Schwinger's 
proper time technique  \cite{Sch51,B2}
and we shall thus address this case in
detail here. It amounts to the evaluation of the vacuum 
permeability:
\begin{equation}\label{form1}
{\cal L}_{\rm eff}(\vec E=0, \vec n \parallel \vec B) =
   -[\Pi(ma)+F_2(ma)]\frac{ \vert B\vert^{\,2}}{2}
    \equiv \frac{-1}{\mu(am)}\frac{ \vert B\vert^{\,2}}{2}\,.
\end{equation}
The magnetic permeability introduced here has
the vacuum limit $\mu\to 1$ for $ma\to \infty$, assured by 
the renormalization process carried out below.
\par
We now turn to determine the effective 
Casimir-Euler-Heisenberg action and $\mu(am)$ in particular. 
So let us consider the quantum vacuum of a Dirac field of 
mass $m$ and charge $e$ in the non-trivial topology of 
$\R$(time)$\times\R^2$(xy-plane)$\times {\rm S}^1$ in the presence 
of the constant uniform magnetic field ${\bf B}$. 
We use Schwinger's proper-time method \cite{Sch51} 
in order to calculate the effective Lagrangian. We start our 
calculations with the proper-time representation for the  
effective action with regularization provided by a cutoff 
$s_o$ in the proper-time $s$:
\begin{equation}\label{calW}
{\cal W}^{(1)}={i\over 2}\int_{s_o}^\infty
{ds\over s}\;Tr\,e^{-isH}\,,
\end{equation}
where $Tr$ stands for the total trace and $H$ is the proper-time 
Hamiltonian, which for the case at hand is given by
$H=(p-eA)^2-(e/2)\sigma_{\mu\nu} F^{\mu\nu}+m^2$,
where $p$ has components $p_\mu=-i\partial_\mu$, $A$ is the 
electromagnetic potential and $F$ is the electromagnetic field, 
which is being contracted with the combination of gamma matrices
$\sigma_{\mu\nu}=$$i[\gamma_\mu,\gamma_\nu]/2$. Using the 
anti-periodic boundary condition we find for $p_z$ 
the eingenvalues $\pm\pi n/2a$ ($n\in 2\Nat-1$), 
where $\Nat$ represents the set of positive integers.
The components which are parallel to the plates are 
constrained into the Landau levels created by the magnetic 
field ${\bf B}$; we call $B$ the component of ${\bf B}$ 
perpendicular to the xy-plane and consider ${\bf B}$ oriented 
in such a way that $eB$ is positive. 
A straightforward calculation yields for the trace in 
Eq.\,(\ref{calW}) the following expression:
\begin{equation}\label{Tr}
Tr\,e^{-isH}=e^{-ism^2}\sum_{\alpha=\pm 1}2 
\sum_{n\in 2\Nat-1}2\,e^{-is(\pi n/2a)^2}
\sum_{n^\prime \in \Nat-1}{eB\ell^2\over 2\pi} 
                          e^{-iseB(2n^\prime+1-\alpha)}
\int{dt\,d\omega\over 2\pi}e^{is\omega^2}\,,
\end{equation}
where the first sum is due to the four components of the Dirac spinor,
the second sum is over the eigenvalues obtained from the anti-periodic
boundary condition, the third sum comes from the Landau levels 
with the corresponding multiplicity factor due to  degeneracy, 
and the integration is done over the observation time $T$ 
and the continuum of eigenvalues $\omega$ of the
operator $p^o$. Following Schwinger \cite{Sch92}
we now use Poisson's formula \cite{Poisson} to rewrite the second
sum in  Eq.\,(\ref{Tr}) in the following form:
\begin{equation}
\sum_{n\in 2\Nat-1}2\, e^{-is(n\pi/2a)^2}={2a\over 2\sqrt{i\pi\,s}}\,
\left[1+2\sum_{n\in \Nat}(-1)^n e^{i(an)^2/s}\right]\,,
\end{equation}
The sum over the Landau levels is trivially obtained in terms of 
hyperbolic functions in such a way that the trace in (\ref{Tr})
takes the form:
\begin{equation}
Tr\,e^{-isH}={T\,2a\ell^2\over 4\pi^2i}{e^{-ism^2}\over s^2}
[1+2\sum_{n=1}^{\infty}(-1)^n e^{i(an)^2/s}][1+iseB\,L(iseB)]\,,
\end{equation}
where $L(\xi)=\coth \xi-\xi^{-1}$ is the Langevin function.
Substituting this expression into equation Eq.\,(\ref{calW}) we obtain 
the effective action as
\begin{equation}
{\cal W}^{(1)}=[{\cal L}_{HE}^{(1)}(B)+{\cal L}_{HEC}^{(1)}(B,a)]
              \,T2a\ell^2\,,
\end{equation}
where
\begin{equation}\label{HE}
{\cal L}^{(1)}_{HE}(B)={1\over 8\pi^2}\int_{s_o}^\infty {ds\over s^3}
e^{-ism^2}\,[1+iseB\,L(iseB)]
\end{equation}
is the Euler-Heisenberg contribution to the effective Lagrangian and 
\begin{equation}\label{HEC}
{\cal L}_{HEC}^{(1)}(B,a)=\sum_{n=0}^{\infty} {(-1)^n\over 4\pi^2}
\int_{s_o}^\infty{ds\over s^3}\,e^{-ism^2+i(an)^2/s}\;
[1+iseB\,L(iseB)]\; ,
\end{equation}
is the  Casimir-Euler-Heisenberg contribution
to the effective Lagrangian. Both 
contributions are in unrenormalized form
and renormalization is required before we can remove
the cutoff $s_o$. 
\par
The usual renormalization condition 
requiring that the observable Maxwell charge
is seen at large distances requires that the compactification 
size $a$ is taken to infinity first. However, 
since  the HEC contribution is
finite for $s_0\to 0$ for any $a$, a careful study of these limits
in the Abelian theory is not necessary, and we can proceed as usual:  
an expansion of $1+$$iseBL(iseB)$ in powers of $eB$ yields the one 
substraction and one 
renomalization constant.  The first, constant, expansion term
can be subtracted from the HE-part of the effective Lagrangian, 
as it is not field dependent. In the limit 
$s_o\rightarrow 0$ this constant vacuum action 
due to free fermi fluctuations tends to $-m^4\Gamma(-2)/8\pi^2$,
where $\Gamma$ is the Euler gamma function. It is generally 
believed that a complete quantum field theory will have equal number of
fermion and boson degrees of freedom, in which case the divergent
component in the zero point contributions to the World action can 
cancel out.
\par
The second term in the expansion  is 
proportional to the Maxwell Lagrangian,
\begin{equation}\label{Maxwell}
{\cal L}^{(0)}(B)=-{1\over 2}B^2\; ,
\end{equation}
with a constant of proportionality $Z_3^{-1}-1=e^2\Gamma(0)/12\pi^2$ 
in the limit $s_o\rightarrow 0$. We absorb  this constant into Maxwell 
Lagrangian by a renormalization of $B$:
define the renormalized field as $B_R=BZ_3^{-1/2}$ and the renormalized 
charge as $e_R=eZ_3^{1/2}$. After subtractions and conversions to these 
renormalized quantities, ${\cal L}^{(1)}_{HE}$ in Eq.\,(\ref{HE}) 
becomes free of spurious terms and is well-defined in the limit 
$s_o\rightarrow 0$ while in the Maxwell Lagrangian Eq.\,(\ref{Maxwell}), 
the bare field is replaced by $B_R$. The HEC-component ${\cal L}^{(1)}_{HEC}$ 
in Eq.\,(\ref{HEC}) depends on $e$ and $B$ only through the
product $eB= e_RB_R$ and thus it does not change in form. Since
it is finite and vanishes for $a\to \infty$ no further renormalization is 
required to render ${\cal L}_{HEC}^{(1)}$ 
in Eq.\,(\ref{HEC}) well-defined. 
\par
We immediately drop the  subindex 
$R$ indicating renormalization, and assemble the 
renormalized contributions of Eq.\,(\ref{Maxwell}),
Eq.\,(\ref{HE}) and Eq.\,(\ref{HEC}) to write the complete
renormalized effective Lagrangian as:
\begin{eqnarray}\label{L}
{\cal L}(B,a)&=&-{1\over 2}B^2-
{1\over 8\pi^2}\int_{0}^\infty {ds\over s^3}
e^{-sm^2}\,\biggl[seB\,L(seB)-{1\over 3}(seB)^2\biggr]+\nonumber\\
&+&\sum_{n=1}^{\infty} {(-1)^{n-1}\over 4\pi^2}
\int_{0}^\infty{ds\over s^3}\,e^{-sm^2-(an)^2/s}\;
[1+seB\,L(seB)]\,,
\end{eqnarray}
where the integration axis $s$ has been rotated to $-is$  \cite{Sch51} 
and the cutoff $s_o$ has been removed. In the remainder of this 
paper we explore  interesting features of this effective action.
\par
We notice three contributions in Eq.\,(\ref{L}):\\
i) the (renormalized) 
Maxwell Lagrangian ${\cal L}^{(0)}(B)$, given by the quadratic term 
$-B^2/2$;\\
ii) the renormalized  Euler-Heisenberg 
Lagrangian Eq.\,(\ref{HE}):
\begin{equation}\label{HEr}
{\cal L}^{(1)}_{HE}(B)=
 -{1\over 8\pi^2}\int_{0}^\infty {ds\over s^3} 
e^{-sm^2}\biggl[seB\,L(seB)-{1\over 3}(seB)^2\biggr]\,.
\end{equation}
For $B$ small compared to the critical field $B_{\rm cr}=m^2/e$ we can 
expand this Lagrangian in powers of $B^2$ to obtain
\begin{equation}\label{HEserie}
{\cal L}^{(1)}_{HE}=\sum_{k=2}^{\infty}{(-1)^k\over 8\pi^2}
{2^{2k} \vert B_{2k}\vert\; m^4 \over 
        2k(2k-1)(2k-2)}{B^{2k}\over B_{\rm cr}^{2k}} \,,
\end{equation}
where $B_{2k}$ is the $2k$-th Bernoulli number. This expression shows 
that the lowest order contribution from the (renormalized)
Euler-Heisenberg Lagrangian 
to Maxwell Lagrangian is a term in $B^4$.\\
iii) The third contribution in Eq.\,(\ref{L}) 
is given by the Casimir-Euler-Heisenberg Lagrangian 
${\cal L}^{(1)}_{HEC}$ in Eq.\,(\ref{HEC}) with renormalized charge
and field.
\par
We are in particular interested here in any modification of
the magnetic permeability of the vacuum, $\mu(B,am)$ 
as defined by the derivative of the complete effective 
Lagrangian Eq.\,(\ref{L}) with respect to $-B^2/2$:
\begin{eqnarray}\label{mub}
H\equiv -\frac{\partial {\cal L}}{\partial B}
  \equiv \frac{1}{\mu(B,am)} B\,.
\end{eqnarray}
By using Eq.\,(\ref{HEserie})
and employing the  formula {\bf 3.471},9 in Ref.\,\cite{GR}  we obtain:
\begin{eqnarray}\label{mufunction}
\frac{1}{\mu(B,am)}&=&\frac{1}{\mu(am)}-\\
\hspace*{-0.3cm}&-&\hspace*{-0.3cm}\sum_{k=2}^\infty\biggl[ 1 -
\sum_{n=1}^\infty (-1)^{n-1}
{2^2(amn)^{2k-2}\over(2k-3)!} K_{2k-2}(2amn) \biggr]
{(-1)^k\over 8\pi^2}
{2^{2k} \vert B_{2k}\vert e^2\over(2k-1)(2k-2)}
{B^{2k-2}\over B_{cr}^{2k-2}}\,,\nonumber
\end{eqnarray}
where we introduced
\begin{equation}\label{mu}
{1 \over \mu(B\to 0,am)}\equiv{1 \over \mu(am)}=
1-{e^2\over 3\pi^2}\sum_{n=1}^{\infty}(-1)^{n-1}
K_o(2amn)\,.
\end{equation}
In the weak field regime the vacuum  
permeability  Eq.\,(\ref{mu})  is the dominant term
in the expansion given by Eq.\,(\ref{mufunction}). 
\par
The corresponding separation of the
Casimir-Euler-Heisenberg term in Eq.\,(\ref{HEC}) is:
\begin{equation}\label{HECr}
{\cal L}^{(1)}_{HEC}(B)={(am)^2\over 2\pi^2 a^4} 
\sum_{n=1}^{\infty} {(-1)^{n-1}\over n^2}K_2(2amn)+{1\over 2}
\biggl[1-{1\over \mu(am)}\biggr]B^2+{\cal L}_{HE}^{(1)\prime}(a,B) \,, 
\end{equation}
with
\begin{equation}\label{HE'}
{\cal L}_{HE}^{(1)\prime}(a,B)=\sum_{n=1}^{\infty}{(-1)^{n-1}\over 4\pi^2}
\int_{0}^\infty {ds\over s^3} 
            e^{-sm^2-(an)^2/s}\,[seB\,L(seB)-{1\over 3}(seB)^2]\,.
\end{equation}
The first term in Eq.\,(\ref{HECr}) is the negative of the Casimir energy 
density of the fermionic field with anti-periodic boundary condition; 
it has no importance in the present discussion because it is independent of $B$.
\par
\begin{figure}[tb]
\begin{center}
%
\hspace*{0.6cm}\psfig{width=15cm,figure=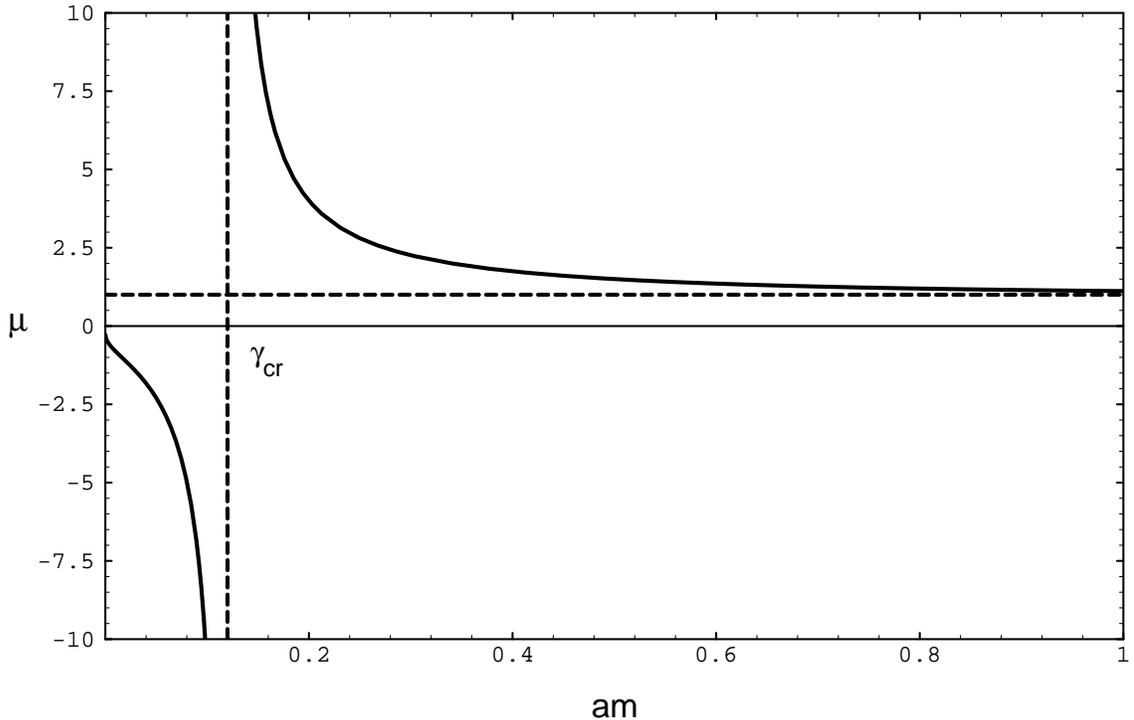}
\end{center}
\vskip -0.9cm
\protect\caption{Permeability of Fermionic vacuum as a 
function of $am$ for $\alpha=3\pi/4$.}
\protect\label{muvam}
\end{figure}

We illustrate the behavior of the vacuum permeability 
$\mu(am)$, Eq.\,(\ref{mu}), as a function of $am$ in figure \ref{muvam}. 
We see that it reduces 
to 1 when the compactification disappears ($a\rightarrow\infty$). 
In the general case $a<\infty$ the permeability of the Fermionic 
vacuum confined in $\R^2\times {\rm S}^1$ is determined by 
the series of Bessel 
functions in Eq.\,(\ref{mu}). From the properties of Bessel 
functions \cite{GR} it is easy to see that the behavior obtained 
numerically in figure \ref{muvam} taking $\alpha=3\pi/4$ for
illustrative purposes, is correct. There is a 
value $\gamma_{\rm cr}\equiv ma_{\rm cr}$  at which a transition between 
paramagnetic and  diamagnetic permeability occurs.
For $am$$<\gamma_{\rm cr}$ we have a diamagnetic permeability and for 
$am$$>\gamma_{\rm cr}$ we have a paramagnetic permeability. From
 Eq.\,(\ref{mu}) we obtain this critical value $\gamma_{\rm cr}$ 
as defined by the equality
\begin{equation}\label{gammacr}
\sum_{n=1}^{\infty}(-1)^{n-1} K_o(2ma_{\rm cr}n)= {3\pi\over 4\alpha}\,,
\end{equation}
where it appears the fine structure constant of the Dirac particle,
$\alpha=$$e^2/4\pi$, which is the magnitude that determines  $a_{\rm cr}$. 
Typical values of the Bessel function 
$K_o$ show that $a_{\rm cr}$ is extremely small for the usual 
values of $\alpha$ in the case of electrons and quarks. In this case it is 
easy to find the estimation (cf. formula {\bf 8.526.2} in  \cite{GR}) 
\begin{equation}\nonumber
ma_{\rm cr}\simeq {\pi\over 2\, e^C} e^{-3\pi/2\alpha}\,,
\end{equation}
where $C$ is the Euler constant. This estimation shows that 
$ma_{\rm cr}$ is exceedingly small for the cases under 
consideration (for the electron its order of magnitude 
is $10^{-282}$); in those cases the vertical asymptote 
in Figure 1 becomes the vertical axis of the graph, 
which is reduced to the curve at the right of this 
asymptote. Therefore, we are more than justified in 
attributing paramagnetic properties to the Fermionic 
vacuum under consideration.
%
%
%
%
On the other hand it is reassuring 
to obtain from Eq.\,(\ref{mu}) a sensible result
even in the limiting situation of $a<< a_{\rm cr}$.
In fact, for extremely dense packed vacuum fluctuations, intuition 
favors diamagnetism as the dominant effect.
\par
Clearly the fermionic vacuum 
permeability arising in  typical 
physical sizes involving electromagnetism is very small.
If we take for the electron vacuum field $a$ in the range of
nanometer the change in permeability given by Eq.\,(\ref{mu}) 
can be taken as zero (it is less than $10^{-2200}$). On the 
other hand if we consider the vacuum field of $u$, $d$ and $s$ 
quarks inside a hadron, assuming that our expression derived for 
confinement in $\R^2\times {\rm S}^1$ provides a magnitude
estimate for quarks confined in the region inside the spherical 
surface ${\rm S}^2$ (and assuming the
radius of ${\rm S}^1$ and ${\rm S}^2$ are roughly the same). Numerically 
obtained changes in permeability are of magnitude 
$10^{-3}$, $10^{-3}$ and $10^{-4}$, respectively, 
for $u$, $d$ and $s$ quarks and $a\approx 1\,$\,fm. 
This effect falls in the range
of typical material constants: for example  aluminum, for which 
$\mu-1$$\approx 2.3\times 10^{-4}$ at $20^o$C.
These  changes in vacuum 
permeability show that equation Eq.\,(\ref{mu}) yields no 
contribution to QED at realistic domain scale. However, these formal
results are of potential importance to the understanding of the QED vacuum 
inside a hadron. Moreover, the transition point from paramegnetic 
to diamagnetic  vacuum deserves special attention in the 
studies of dimensional compactifcation. 
We further observe that  our study can be
adapted to the more complicated gauge groups 
and be applied to the study of QCD in particular, where the 
coupling constant is large, and thus color magnetic permeability
(and permitivity as well)
is expected to be quite different from the perturbative
vacuum value within the confinement volume.
\par
We have obtained  by an explicit calculation the magnetic
permeability of the QED vacuum in the infrared limit, including 
the dependance on the size of confining regions.
The effect is found to be exceedingly small and should not 
influence  the birefringence  experimental tests of the 
Euler-Heisenberg effective Lagrangian. On the other hand
both QED and QCD effects within the hadron de-confinement 
region promise to be of interest and we hope to return
to study these effects in near future.
\subsection*{Acknowledgments:}
J.R. would like to thank E. Zavattini and C. Rizzo for 
stimulating discussions and continued interest in this work.
M. V. C.-P. and C. F. would like to acknowledge CNPq 
(The National Research Council of Brazil) 
for partial financial support. 
J.R. acknowledges partial support 
by  DOE, grant DE-FG03-95ER40937\,.
\vfill\eject

\end{document}